\documentclass{biophys-new}
\usepackage[utf8]{inputenc}
\usepackage{graphicx}
\usepackage[colorlinks,allcolors=cyan!70!black]{hyperref}

\usepackage{lipsum}

\title{Cell migration: Beyond Brownian motion}
\runningtitle{Biophysical Journal Template} 

\author[1,2,*]{Rainer Klages}
\runningauthor{Rainer Klages} 

\affil[1]{School of Mathematical Sciences, Queen Mary University of London,
Mile End Road, London E1 4NS, UK}

\affil[2]{London Mathematical Laboratory, 8 Margravine Gardens, London W6 8RH, United Kingdom}

\corrauthor[*]{r.klages@qmul.ac.uk}

\papertype{New and Notable}

\begin{document}

\begin{frontmatter}

\end{frontmatter}

\begin{figure}[hbt!]
\centering
\includegraphics[width=0.5\linewidth]{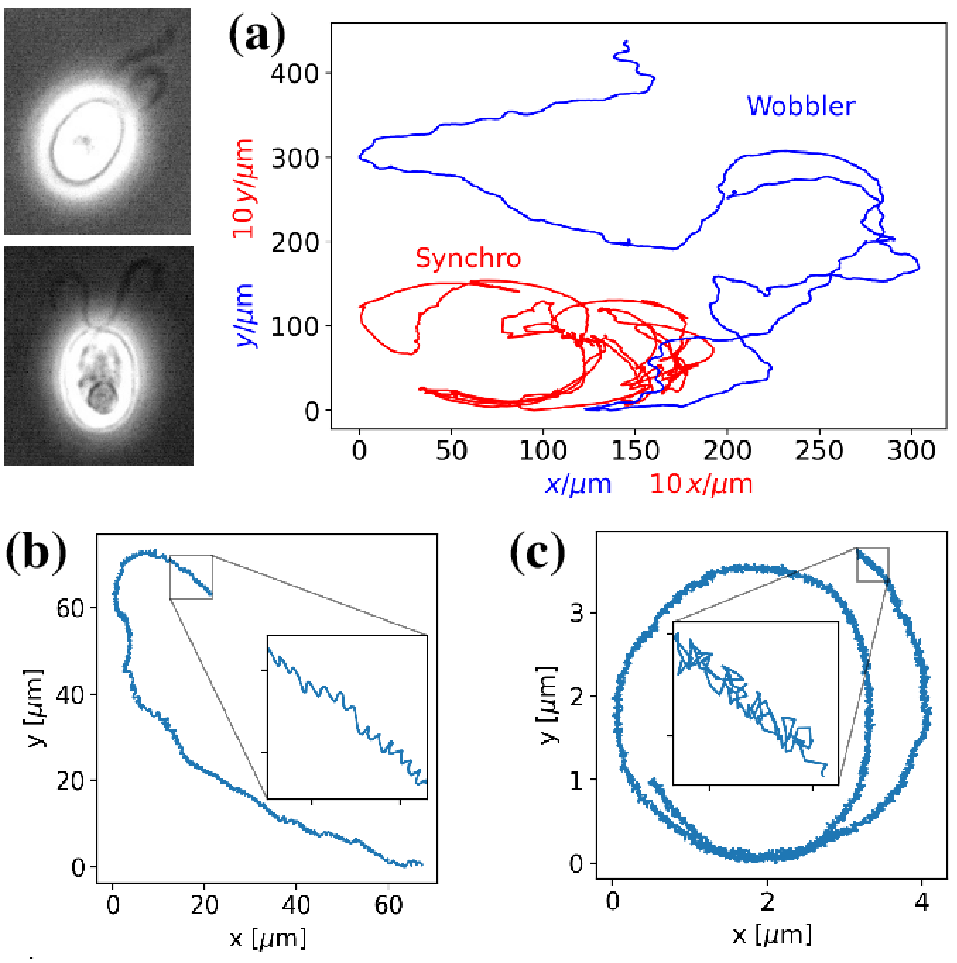}
\caption{(a) Phase-contrast microscopy images and swimming paths of
  {\em Chlamydomonas reinhardtii} algae. These cells exhibit two different
  swimming modes depending on their diameter, characterised by either
  wobbler (top image) or synchro-type (bottom) flagellar motion. Shown
  to the right of (a) are sample paths up to a minute, which on these
  scales look quite `random'. (b), (c) Higher resolution data up to a
  few seconds, plus blow-ups, revealing a non-trivial complex fine
  structure of the swimming paths for (b) wobbler and (c) synchro
  (adapted from \cite{KMBSN24}, with kind permission by the authors).}
\label{fig:cellpaths}
\end{figure}

Cell migration is vital, fulfilling essential biological tasks like
morphogenesis, wound healing, or the killing of pathogens in
organisms. But cell migration also drives detrimental processes like
tumor metastasis or inflammation reactions. To understand cell
migration by classifying different types of cells defines a
fundamental problem of cell science. Recording under a microscope the
paths of single cells reveals random-looking migration (see
Fig.~\ref{fig:cellpaths}(a), right) that, in terms of physics, reminds
of the diffusive Brownian motion of a tracer particle in a fluid. This
tracer motion can be described by the Langevin equation \cite{Lang08},
a stochastic version of Newton's second law, as it models the random
collisions of a tracer with the fluid molecules by a random
force. Around the same time as Langevin published his famous equation,
Pearson proposed to model the movements of organisms by random walks
\cite{Pea06}. This idea stipulates that an organism moves in a random
direction over a certain distance per time step. These two approaches
coined a paradigm that persisted for almost a century, suggesting that
biological movements are in general `so random' that they can be
modelled by stochastic processes {\em without memory}. Pearson
conveyed this idea by the picture of a drunken sailor, which can be
found in many textbooks of statistical mechanics. That is, an agent
makes the next movement step by not remembering the previous one. In
mathematical terms, this complete loss of memory is called a {\em
  Markov property}.

Analysing experimentally recorded paths of different types of
migrating cells, decades ago it was reported that cells move according
to the predictions of random walks \cite{GaBo70}, respectively
Langevin dynamics \cite{DuBr87}. Subsequent work, however, challenged
this simple picture by revealing the existence of memory in terms of
non-Markovian correlations between different time steps
\cite{HLCC94}. Especially over the past two decades, many articles
appeared suggesting that cell migration is typically way more complex
than Markovian random motion
\cite{RUGOS00,DKPS08,HaBa12,MSDRN20}. There are indeed many physical
and biological reasons to doubt a simplifying description of cell
migration patterns by Brownian motion. First of all, a Brownian tracer
in a fluid diffuses {\em passively}, driven by the collisions with the
surrounding particles, while a cell moves {\em actively} by itself
converting chemical into kinetic energy \cite{RBELS12}. Active
biological motion, in turn, involves complex microscopic processes,
like cytoskeletal fluctuations driving cell crawling, or the beating
of flagella steering microswimmers. On top of this, there are
interactions with the environment via surface adhesion and biochemical
cell-signalling cascades. {\em Per se} there is no reason why all
these complex processes across different spatio-temporal scales should
integrate out to yield purely Markovian dynamics. However, assessing
non-Markovian correlations of cell migration in experiments for
constructing data-driven mathematical models reproducing memory is a
notoriously difficult task. It necessitates to compute correlation
functions, related to what statisticians call covariances, which are
not easy to reliably extract from data, nor to analyse.

In their new contribution to {\em Biophys.J.} \cite{KMBSN24} Klimek et
al.\ tackle this problem by analysing experimental trajectories of
unicellular {\em Chlamydomonas reinhardtii} microalgae. These are
microswimmers generating different movement paths, depending on
whether the flagella driving their motion move synchronously or not,
see Fig.~\ref{fig:cellpaths}(b),(c). Using a specific experimental
set-up by which they constrain the microswimmers between two glass
plates so that they do not move out of focus of the microscope, they
generated 3000 position data points for 59 algae each. This data is
analysed by using a novel form of generalised Langevin dynamics
incorporating non-Markovian memory, thus significantly going beyond
Brownian motion.

Very interestingly, the key feature of their theory, the generalised
Langevin equation (GLE), can be derived from first principles. Let us
assume that all particles forming a system interact according to
Newton's classical mechanical equations of motion. Let us identify one
of them, with a bigger mass than the others, as a `walker' while the
others act as a `driving' molecular bath. It is well-known that for
this setting, applying a (Mori-Zwanzig) projection operator formalism
by making simplifying assumptions, in particular assuming complete
loss of memory, the Markovian Langevin equation modelling Brownian
motion can be derived. However, this approach can be generalised by
including memory in the dynamics leading to non-Markovian GLEs
\cite{Zwan01}. Very recently, Netz and colleagues have worked out a
new, more detailed theory along these lines leading to a novel type of
GLE \cite{ASDN22}. This theory has already been applied to reveal
non-Markovian effects in protein folding \cite{DAKKTN23} and human
breast cancer cell migration \cite{MSDRN20}. Reference~\cite{KMBSN24}
continues this line of research by providing compelling evidence that
their framework can be used to construct a non-Markovian GLE from data
for modelling the motion of {\em C.~reinhardtii}
microswimmers. Combining this method with an unbiased cluster analysis
they show that, without any prior knowledge, different types of cells
can be distinguished by purely analysing their migration
patterns. Their numerical findings perfectly match to the different
types of cells identified by visual inspection, cf.\ the different
paths in Fig.~\ref{fig:cellpaths} (b),(c). This paves the way for
devising a general method to classify different cells based on motion
trajectories only.

A fundamental open question is the generality, or perhaps even
universality, of both their underlying theory and their approach for
experimental data analysis. One may challenge this framework along
three lines: First, the GLE devised by Netz et al.\ is by far not the
only approach to model non-Markovian stochastic dynamics. Apart from
many other types of GLEs, like fractional, scaled or superstatistical
Brownian motion, there is the very different class of sub- or
superdiffusive continuous time random walks \cite{KRS08,MJCB14}. The
Gaussian noise generating Brownian motion-type dynamics may
furthermore be replaced by power laws yielding L\'evy processes
\cite{VLRS11}. The virtue of the theory by Netz et al.\ is that, in
contrast to all these other stochastic processes, it has a sound
physical foundation in terms of classical mechanics. On the other
hand, one may wonder whether all physical and biological systems, no
matter how complex they are, need to be understood starting from
microscopic Hamiltonian many-particle dynamics. Furthermore,
non-Markovian dynamics driven by power laws exhibits intricate
phenomena like non-stationarity, ageing, weak ergodicity breaking and
long-term anomalous diffusion, all well observed in physical and
biological experiments \cite{KRS08,MJCB14}. To what extent the
theoretical framework put forward by Netz et al.\ can cover these
properties as well remains an interesting open question.

Second, this theory relies on the important assumption of Gaussianity
(normality) of relevant probability distributions by exploiting
fluctuation-dissipation relations (FDRs). But the former excludes
L\'evy walks that attracted considerable attention over the past two
decades for modelling organismic movement \cite{VLRS11}, including
cell migration \cite{HaBa12}. Since L\'evy distributions obey a
generalised central limit theorem \cite{KRS08}, there is a reason why
one should expect to see them in nature. Another interesting point is
that for describing {\em C.~reinhardtii} microswimming the authors map
a non-Markovian GLE without FDR onto an effective GLE with FDR. But
this again relies on Gaussianity. Whether this method can be extended
to non-Gaussian dynamics, and to non-equilibrium processes like
chemotaxis, poses further inspiring open questions \cite{Netz23}.

Third, the validity of FDRs is intimately related to distinguishing
active from passive dynamics. As explained at the beginnng of this
article, the original Langevin equation models the passive motion of a
Brownian tracer. This follows from its physical derivation based on
microscopic Hamiltonian dynamics yielding diffusion in non-living
matter. However, within the past two decades the new concept of active
matter was developed to understand biological processes in living
systems, like the formation of tissues, swarming, or crowd dynamics. A
special case within this framework is the theory of active Brownian
particles, modelling the movements of small agents like bacteria,
daphnia, or artificial Janus particles \cite{RBELS12,BeDiL16}. All
active Brownian particle models feature broken FDRs, mathematically
expressing the externally driven activity of biological dynamics. On
the other hand, they have all been formulated phenomenologically,
instead of being derived from Hamiltonian dynamics, like GLEs
exhibiting FDRs. To what extent standard models of active Brownian
motion \cite{RBELS12,BeDiL16} can be derived from first principles
forms yet another crucial open question. Recently, the dynamics of a
passive tracer in an active bath of microswimmers has been calculated
from microscopic non-Hamiltonian equations of motion \cite{KSCB20},
taking special care of the hydrodynamic interaction between tracer and
bath. Combining this with other techniques for constructing a
biomolecular model of cell migration poses a formidable challenge for
future research.

\section*{Declaration of interests}
The author declares no competing interests.



\begin{thebibliography}{20}
\providecommand{\url}[1]{\texttt{#1}}
\providecommand{\urlprefix}{ }

\bibitem[Langevin(1908)]{Lang08}
Langevin, P., 1908.
\newblock Sur la th\'eorie du mouvement brownien.
\newblock \emph{C.R. Acad. Sci. (Paris)} 146:530--533.
\newblock See also the English translation in Am. J. Phys. {\bf 65}, 1079
  (1997).

\bibitem[Pearson(1906)]{Pea06}
Pearson, K., 1906.
\newblock Mathematical contributions to the theory of evolution - A
  mathematical theory of random migration.
\newblock \emph{Biometric ser.} 3:54.

\bibitem[Klimek et~al.(2024)Klimek, Mondal, Block, Sharma, and Netz]{KMBSN24}
Klimek, A., D.~Mondal, S.~Block, P.~Sharma, and R.~R. Netz, 2024.
\newblock Data-driven classification of individual cells by their non-Markovian
  motion.
\newblock \emph{Biophys. J.} 123:1173--1183

\bibitem[Gail and Boone(1970)]{GaBo70}
Gail, M., and C.~Boone, 1970.
\newblock The locomotion of mouse fibroblasts in tissue culture.
\newblock \emph{Biophys. J.} 10:980--993.

\bibitem[Dunn and Brown(1987)]{DuBr87}
Dunn, G., and A.~Brown, 1987.
\newblock A unified approach to analysing cell motility.
\newblock \emph{J. Cell Sci. Suppl.} 8:81--102.

\bibitem[Hartmann et~al.(1994)Hartmann, Lau, Chou, and Coates]{HLCC94}
Hartmann, R., K.~Lau, W.~Chou, and T.~Coates, 1994.
\newblock The fundamental motor of the human neutrophil is not random: Evidence
  for local non-Markov movement in neutrophils.
\newblock \emph{Biophys. J.} 67:2535--2545.

\bibitem[Rieu et~al.(2000)Rieu, Upadhyaya, Glazier, Ouchi, and Sawada]{RUGOS00}
Rieu, J., A.~Upadhyaya, J.~Glazier, N.~Ouchi, and Y.~Sawada, 2000.
\newblock Diffusion and deformations of single Hydra cells in cellular
  aggregates.
\newblock \emph{Biophys. J.} 79:1903--1914.

\bibitem[Dieterich et~al.(2008)Dieterich, Klages, Preuss, and Schwab]{DKPS08}
Dieterich, P., R.~Klages, R.~Preuss, and A.~Schwab, 2008.
\newblock Anomalous dynamics of cell migration.
\newblock \emph{PNAS} 105:459--463.

\bibitem[Harris et~al.(2012)Harris, Banigan, Christian, et~al.]{HaBa12}
Harris, T., E.~Banigan, D.~Christian, et~al., 2012.
\newblock Generalized {L}\'{e}vy walks and the role of chemokines in migration
  of effector CD8{+} T cells.
\newblock \emph{Nature} 486:545--548.

\bibitem[Mitterwallner et~al.(2020)Mitterwallner, Schreiber, Daldrop, R\"adler,
  and Netz]{MSDRN20}
Mitterwallner, B.~G., C.~Schreiber, J.~O. Daldrop, J.~O. R\"adler, and R.~R.
  Netz, 2020.
\newblock Non-Markovian data-driven modeling of single-cell motility.
\newblock \emph{Phys. Rev. E} 101:032408.

\bibitem[Romanczuk et~al.(2012)Romanczuk, B{\"a}r, Ebeling, Lindner, and
  Schimansky-Geier]{RBELS12}
Romanczuk, P., M.~B{\"a}r, W.~Ebeling, B.~Lindner, and L.~Schimansky-Geier,
  2012.
\newblock Active Brownian particles.
\newblock \emph{Eur. Phys. J. Spec. Top.} 202:1--162.

\bibitem[Zwanzig(2001)]{Zwan01}
Zwanzig, R., 2001.
\newblock Nonequilibrium statistical mechanics.
\newblock Oxford University Press, Oxford.

\bibitem[Ayaz et~al.(2022)Ayaz, Scalfi, Dalton, and Netz]{ASDN22}
Ayaz, C., L.~Scalfi, B.~A. Dalton, and R.~R. Netz, 2022.
\newblock Generalized Langevin equation with a nonlinear potential of mean
  force and nonlinear memory friction from a hybrid projection scheme.
\newblock \emph{Phys. Rev. E} 105:054138.

\bibitem[Dalton et~al.(2023)Dalton, Ayaz, Kiefer, Klimek, Tepper, and
  Netz]{DAKKTN23}
Dalton, B.~A., C.~Ayaz, H.~Kiefer, A.~Klimek, L.~Tepper, and R.~R. Netz, 2023.
\newblock Fast protein folding is governed by memory-dependent friction.
\newblock \emph{PNAS} 120:e2220068120.

\bibitem[Klages et~al.(2008)Klages, Radons, and Sokolov]{KRS08}
Klages, R., G.~Radons, and I.~M. Sokolov, editors, 2008.
\newblock Anomalous transport: Foundations and Applications.
\newblock Wiley-VCH, Berlin.

\bibitem[Metzler et~al.(2014)Metzler, Jeon, Cherstvy, and Barkai]{MJCB14}
Metzler, R., J.-H. Jeon, A.~G. Cherstvy, and E.~Barkai, 2014.
\newblock Anomalous diffusion models and their properties: non-stationarity{,}
  non-ergodicity{,} and ageing at the centenary of single particle tracking.
\newblock \emph{Phys. Chem. Chem. Phys.} 16:24128--24164.

\bibitem[Viswanathan et~al.(2011)Viswanathan, {da Luz}, Raposo, and
  Stanley]{VLRS11}
Viswanathan, G., M.~{da Luz}, E.~Raposo, and H.~Stanley, 2011.
\newblock The Physics of Foraging.
\newblock Cambridge University Press, Cambridge.

\bibitem[Netz(2023)]{Netz23}
Netz, R., 2023.
\newblock Derivation of the non-equilibrium generalized Langevin equation from
  a generic time-dependent Hamiltonian.
\newblock \emph{preprint arXiv:2310.00748} .

\bibitem[Bechinger et~al.(2016)Bechinger, Di~Leonardo, L\"owen, Reichhardt,
  Volpe, and Volpe]{BeDiL16}
Bechinger, C., R.~Di~Leonardo, H.~L\"owen, C.~Reichhardt, G.~Volpe, and
  G.~Volpe, 2016.
\newblock Active particles in complex and crowded environments.
\newblock \emph{Rev. Mod. Phys.} 88:045006.

\bibitem[Kanazawa et~al.(2020)Kanazawa, Sano, Cairoli, and Baule]{KSCB20}
Kanazawa, K., T.~G. Sano, A.~Cairoli, and A.~Baule, 2020.
\newblock Loopy L\'evy flights enhance tracer diffusion in active suspensions.
\newblock \emph{Nature} 120:364–367.

\end{thebibliography}

\end{document}